\PassOptionsToPackage{dvipsnames}{xcolor}
\documentclass[11pt,letterpaper]{style}

\usepackage[numbers]{natbib}
\usepackage{graphicx}
\usepackage{booktabs}
\usepackage{amsmath,amsfonts,amssymb}
\usepackage{cleveref}
\usepackage{subcaption}
\usepackage{wrapfig}
\usepackage{multirow}
\usepackage{colortbl}
\usepackage{listings}
\usepackage{xparse}
\usepackage{fontawesome5}
\usepackage{float}
\usepackage{placeins}
\usepackage{threeparttable}

\graphicspath{{./}{fig/}{figures/}{plot/}{pdf/}{table/}}
\usepackage{amsthm}
\usepackage{tcolorbox}
\usepackage{svg}
\tcbuselibrary{skins,breakable}
\tcbuselibrary{listingsutf8}
\usepackage{titletoc}

\usepackage{setspace}
\usepackage{pifont}
\usepackage{mathtools}
\usepackage{enumitem}
\usepackage{arydshln}
\usepackage{bbm}
\usepackage{lineno}
\usepackage{makecell}
\usepackage{adjustbox}
\usepackage{algorithm}
\usepackage{algorithmic}
\usepackage{caption} 
\usepackage{wrapfig}
\usepackage{hyperref}
\usepackage{bbding}
\usepackage{xspace}

\hypersetup{colorlinks=true,linkcolor=red,urlcolor=blue,citecolor={blue}}

\definecolor{mygray}{gray}{0.9}
\definecolor{syncol}{RGB}{243,246,249}
\definecolor{wildcol}{RGB}{215,240,235}
\definecolor{drop1}{RGB}{180,225,220}
\definecolor{drop2}{RGB}{150,210,200}
\definecolor{drop3}{RGB}{120,195,185}
\definecolor{drop4}{RGB}{95,180,170}
\definecolor{drop5}{RGB}{65,160,150}
\definecolor{lightblue}{RGB}{26,82,249}

\newcommand{\tablebodyfont}{\normalfont}

\definecolor{myblue1}{HTML}{0171DC}
\definecolor{myblue2}{HTML}{013978}

\NewDocumentEnvironment{minted}{O{} m +b}{%
}{}

\newcommand{\equal}{\textsuperscript{*}}     

\renewcommand\Authfont{\centering\normalfont\bfseries\fontsize{11}{15}\selectfont}
\renewcommand\Affilfont{\centering\normalfont\fontsize{10}{15}\selectfont}

\title{MAVE: A Multi-Agent Framework for Grounded Pedagogical Feedback from Online Class Videos}
\runningtitle{MAVE: A Multi-Agent Framework for Grounded Pedagogical Feedback from Online Class Videos}

\author{%
    {\Authfont
    \textbf{Qingyue Chen}\textsuperscript{1} \quad
    \textbf{Huaiyuan Yao}\textsuperscript{1} \quad
    \textbf{Wanpeng Xu}\textsuperscript{1} \quad 
    \textbf{Varun Shourie}\textsuperscript{1} \\
    \textbf{Qianwen Wang}\textsuperscript{2} \quad
    \textbf{Hua Wei}\textsuperscript{1}
    }\\
    {\Affilfont
    \textsuperscript{1} Arizona State University \quad
    \textsuperscript{2} University of Minnesota
    }
}

\begin{document}

\begin{abstract}
Timely pedagogical feedback is difficult to scale in online classes, where long and noisy videos make student responses hard to locate and automated systems often produce generic advice weakly grounded in actual classroom behavior. The key challenge is to connect the instructional goal of a specific lesson segment with temporally aligned multimodal evidence of student responses, and then translate the observed gap into an actionable teaching move. We introduce \textbf{MAVE} (\textbf{M}ulti-\textbf{A}gents for \textbf{V}ideo-based \textbf{E}ducational Feedback), a multi-agent framework for grounded pedagogical feedback from online-class videos. MAVE decomposes feedback generation into four roles: the Curator structures the lesson and infers expected student activity; the Observer extracts aligned speech, visual, and interaction evidence; the Assessor estimates the activity supported by student behavior; and the Advisor generates gap-conditioned teaching recommendations. To evaluate this capability, we introduce MAVEBench, a benchmark designed to measure whether feedback systems can ground diagnoses in multimodal student evidence, and generate actionable pedagogical feedback. MAVEBench contains 173 knowledge-point segments from approximately 20 hours of authentic online-class videos across STEM and Humanities, together with comprehensive simulation. Experiments show that MAVE outperforms direct prompting, Reflexion, and TextGrad in diagnostic correctness, instructional alignment, and actionability. 
Our code is available at \url{https://github.com/DaRL-GenAI/MAVE}

\end{abstract}

\newcommand{\TitleLinks}{%
\centering
    \vspace{8pt}
}
\maketitle

\section{Introduction}

Online instruction increasingly takes place in screen-sharing-heavy classrooms, where instructors present materials while student responses are distributed across compressed video tiles, chat, and other fragmented channels. This distribution makes it difficult to monitor how students respond to instructional activities in real time~\cite{rasiah2020business}. In such settings, student responses are often delayed, muted, or only partially observable, making it difficult to distinguish passive listening from productive but hidden activity. Although expert supervisors can convert classroom observations into actionable pedagogical feedback~\cite{fisher2025impact}, such observation does not scale across large courses.

\begin{wrapfigure}{r}{0.48\columnwidth}
  \centering
  \vspace{-8pt}
  \includegraphics[width=0.46\columnwidth]{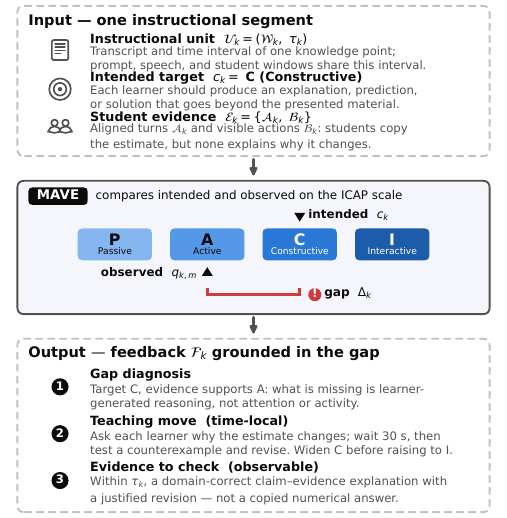}
  \caption{\textbf{Illustrative MAVE input--output transformation.}
  For one instructional segment, MAVE places the intended ICAP target
  and the state supported by aligned student evidence on the ordinal
  ICAP scale, measures the intended and observed gap, and returns a
  time-local teaching move with an observable success criterion.
  Arrows indicate data flow rather than causal effects.}
  \label{fig:model_io_overview}
  \vspace{-8pt}
\end{wrapfigure}

As a result, instructors receive little interpretable evidence about when student engagement breaks down, why specific instructional activities fail, or how interaction patterns could be improved~\cite{gorbunova2026barriers, yao-etal-2026-instructional}; instructional refinement remains anecdotal and cannot be replicated across cohorts. We therefore study the task of generating instructor-facing pedagogical feedback from online-class videos. As illustrated in Figure~\ref{fig:model_io_overview}, the input is a local instructional segment with aligned student-side evidence, and the output is a teaching-oriented diagnosis and concrete instructional move grounded in that evidence. Useful automation must therefore go beyond a single global engagement score: it should identify and interpret behaviors within long videos, relate them to the instructional goal of the corresponding segment, and tie each recommendation to the student evidence behind it.

\begin{figure*}[t]
  \centering
  \begin{subfigure}[b]{0.33\textwidth}
    \centering
    \includegraphics[width=\linewidth]{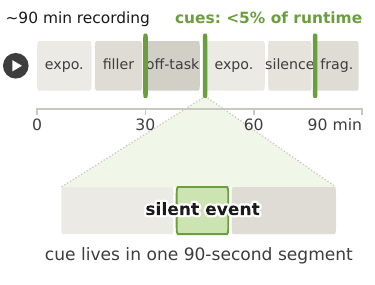}
    \caption{Long, noisy videos.}
    \label{fig:challenge_noise}
  \end{subfigure}
  \hfill
  \begin{subfigure}[b]{0.3\textwidth}
    \centering
    \includegraphics[width=\linewidth]{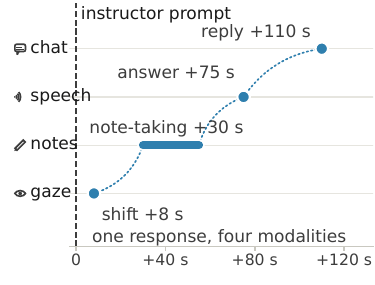}
    \caption{Scattered multimodal evidence.}
    \label{fig:challenge_evidence}
  \end{subfigure}
  \hfill
  \begin{subfigure}[b]{0.33\textwidth}
    \centering
    \includegraphics[width=\linewidth]{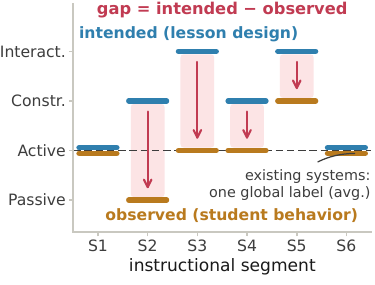}
    \caption{Unmeasured feedback gap.}
    \label{fig:challenge_gap}
  \end{subfigure}
  \caption{Three challenges of generating pedagogical feedback from online classes. (a)~Classroom recordings are long and noisy, so engagement cues occupy a small fraction of the runtime and live inside individual instructional segments. (b)~A single student response is scattered across modalities (gaze, notes, speech, chat) and arrives with delays of seconds to minutes after the instructor's prompt. (c)~Existing systems collapse a session into one global engagement label and never compare the intended cognitive demand of each segment with the engagement students actually exhibit.}
  \label{fig:challenges}
\end{figure*}

Meeting this requirement is hard for three reasons, as shown in Fig.~\ref{fig:challenges}. First, classroom videos are long and noisy: extended exposition, off-task movement, filler speech, and fragmented participation make engagement cues sparse and segment-local~\cite{nagrani2025minerva}. Second, student engagement is multimodal and temporally distributed: note-taking, delayed verbal responses, gaze shifts, and discussion occur asynchronously across modalities. Third, the feedback gap goes unmeasured: existing engagement-analysis systems collapse each segment to a single global label or affect category, and LLM-based feedback approaches rely on unconstrained self-reflection~\cite{shinn2023reflexion,yuksekgonul2024textgrad} or rubric scoring~\cite{chen2024mllm,hong2026rulers}, so neither compares what the lesson aimed to elicit with what students actually showed. Current systems therefore cannot tell whether low engagement reflects expected passive learning, missing instructional scaffolding, or a mismatch between intent and response.

Our central premise is that pedagogical feedback must be grounded in real student responses. LLM judges and rubric scores can help evaluate feedback quality, but they should not be the sole basis for generating feedback. Otherwise, a system may learn to produce recommendations that satisfy rubric language while remaining weakly connected to what students actually did, creating a reward-hacking failure mode. For instructor-facing feedback, a recommendation is useful only when it can be traced to observable student behavior and translated into a concrete teaching move. We therefore frame automated pedagogical feedback as a grounded reasoning chain linking \emph{instructional intent}, \emph{observed student evidence}, and \emph{pedagogical action}.To ground student-state modeling in educational theory, we adopt the ICAP framework~\cite{Chi02102014} provides the shared scale for this chain: each segment can be assigned an intended cognitive demand and an observed student-response state as Passive, Active, Constructive, or Interactive. The intended--observed gap makes the problem diagnosable, but diagnosis alone is not enough. Useful feedback must also select an intervention that matches the direction of the gap, cite the evidence behind it, and provide an observable success criterion.


To operationalize this chain, we introduce \textbf{MAVE}, a multi-agent framework for actionable pedagogical feedback from authentic online-class videos. The \textbf{Curator} structures long videos into instructional segments and infers the intended cognitive-engagement target; the \textbf{Observer} extracts time-aligned speech, visual, and interaction evidence; the \textbf{Assessor} estimates the student-response state and the intended--observed gap; and the \textbf{Advisor} converts this gap into an evidence-grounded teaching recommendation. This decomposition addresses the main challenges of online-class feedback: long and noisy videos, scattered multimodal evidence, and the lack of an explicit feedback gap.

\paragraph{Our Core Contributions.} (1) We formulate instructor-facing pedagogical feedback generation as an evidence-grounded task that connects instructional goals, multimodal student responses, and an actionable teaching move. (2) We propose \textbf{MAVE}, a multi-agent framework with four inspectable roles: the Curator structures the lesson and infers the intended cognitive target; the Observer extracts time-aligned multimodal evidence; the Assessor estimates the observed student-response state and resulting gap; and the Advisor generates evidence-grounded recommendations with observable success criteria. (3) We introduce \textbf{MAVEBench}, a benchmark of 173 knowledge-point segments from approximately 20 hours of authentic online-class videos across STEM and Humanities. Experiments show that MAVE outperforms all baselines in diagnostic correctness, instructional alignment, and actionability.

\section{Related Work}
\subsection{AI in Educational Analytics and Feedback}

Automated classroom analysis has used computer vision, audio processing, and multimodal sensing to infer attention and behavior from faces, gaze, posture, physiological signals, and classroom context \cite{whitehill2014faces,blikstein2013multimodal,liu2025classroom}. Engagement is
typically formulated as multiclass classification or regression
\cite{fredricks2004school,monkaresi2016automated}; although ordinal prediction
is well studied elsewhere \cite{gutierrez2015ordinal}, classroom pipelines
rarely use that structure to represent uncertainty. Systems such as
EmotionCues~\cite{zeng2020emotioncues}, Zheng et al.~\cite{zheng2020intelligent},
and Gao et al.~\cite{gao2025motion} provide behavioral or domain-specific
signals, while adaptive learning methods use explicit preferences or scalar
rewards \cite{rafailov2023direct,okubo2022adaptive,yan2024evidence}. These approaches provide useful but isolated engagement signals, without linking them to the instructional purpose. They therefore do not explain whether student responses match what the instructor intended to elicit, or translate that mismatch into an executable teaching move. MAVE addresses this gap by aligning the instructional segment, the student-response window, and the evidence supporting the ICAP label.

\subsection{Multi-Agent LLMs for Feedback Generation}

LLM agents can be composed through role decomposition, debate, and
cross-verification \cite{du2024improving,liang2024encouraging}. Reflexion
\cite{shinn2023reflexion} and TextGrad \cite{yuksekgonul2024textgrad} refine
outputs through self-critique or textual gradients, and educational applications
use LLMs to assess teaching quality and propose instructional suggestions
\cite{emirtekin2025large,yi2025unified,chen2024mllm,hong2026rulers}. Their optimization signal, however, usually comes from critiques or rubric scores, with limited grounding in student responses. MAVE makes temporal-aligned student evidence an explicit intermediate artifact and uses multiple agents to condition feedback on the resulting intent-response gap.

\begin{figure*}[t]
  \centering
  \includegraphics[width=\textwidth]{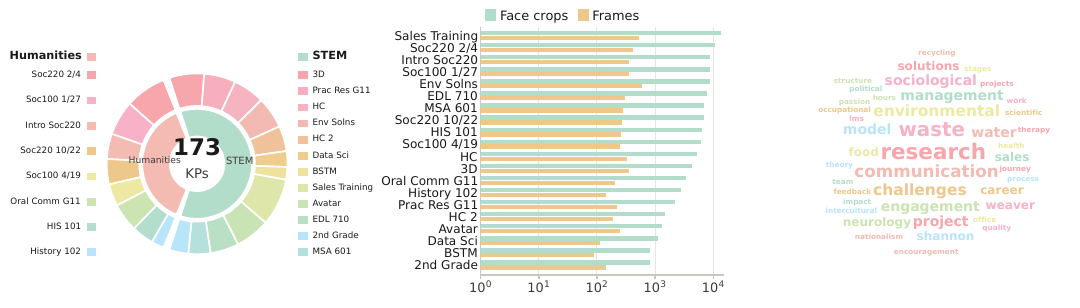}
  \caption{\textbf{MAVEBench overview.} Left: nested composition of the 173
  KPs. Middle:
  per-video face crops and frames after cleaning. Right: topic
  vocabulary.}
  \label{fig:mavebench}
\end{figure*}

\section{MAVE Benchmark}
\label{sec:benchmark}

\subsection{Dataset}
\label{sec:dataset}

To rigorously evaluate MAVE in realistic environments, we introduce \textbf{MAVEBench}—a representative multimodal benchmark bridging authentic classroom dynamics with structured pedagogical assessment. Collected from diverse open-source online lectures, MAVEBench enforces stringent selection criteria: structured pedagogical sequences, recoverable student visual tiles, and clear speech interactions.
The curated benchmark offers broad ecological representativeness, spanning both STEM and Humanities across four distinct educational tiers (K--12, higher education, adult/vocational training, and teacher professional development). This heterogeneous corpus encapsulates diverse teaching formats, noise distributions, and student response modalities. As illustrated in Figure~\ref{fig:mavebench}, MAVEBench exhibits balanced domain coverage, rich multimodal volume, and diverse instructional durations.

\paragraph{Statistics.}
MAVEBench provides 173 KPs totaling 1{,}025 minutes of retained instructional segments, and contains about 5.8K cleaned frames and 110K face crops. The STEM samples contribute 106 KPs, while the Humanities samples contribute 67 KPs. KPs are LLM-segmented and converted into contiguous KP bundles for the controlled transition study. Recoverable class sizes average $\approx$18 students per video (range 6--30), i.e., $\approx$360 student--video instances (students are not re-identified across videos). The long, noisy context and sparse A/C/I behaviors make MAVEBench a challenging testbed for evidence-grounded feedback.

\paragraph{Multimodal preprocessing.}
Each video is transcribed with ASR and segmented by an LLM into semantically
coherent KPs with topics and temporal boundaries; these KPs are the atomic
units for intended/observed ICAP comparison in MAVE.
In parallel, we sample frames, detect Zoom layouts (gallery / sidebar /
spotlight), and crop per-student face windows.
All facial and vocal identifiers are anonymized. Segmentation is constrained to cover the retained instructional timeline with
contiguous, non-overlapping intervals. A boundary is placed at a change in
knowledge objective or instructional activity rather than at a fixed number of
frames. The aligned bundle for KP $k$ contains its transcript, temporal interval,
sampled classroom views, and recoverable student crops. Consequently, all
methods are evaluated on the same evidence bundle and cannot obtain an advantage
from choosing easier boundaries or a different temporal context.

\paragraph{Visual data cleaning.}
Online classroom screenshots contain substantial non-behavioral noise.
We therefore apply a dedicated cleaning stage before engagement analysis:
(i)~\emph{black frames}---empty or near-black student tiles with no visible
person;
(ii)~\emph{icon-only frames}---tiles showing only avatars, profile icons, or
static placeholders without a real face; and
(iii)~\emph{unreadable PPT / slide content}---frames dominated by presentation
slides or screen-share regions whose text/graphics cannot be reliably
recognized as student behavior evidence.
Excluded items are recorded in per-video black/icon/PPT exclusion lists and
removed from the Observer's visual evidence stream, so ICAP assessment is
grounded on observable student activity.

\paragraph{Benchmark unit and quality control.}
The KP is the benchmark unit because it is the shortest interval for which an instructional objective and its
student response can be compared meaningfully. Segment boundaries are checked
for temporal continuity and coverage, and the associated transcript is retained
alongside sampled student views. Preprocessing records why evidence is removed,
so an excluded avatar or slide is not silently interpreted as disengagement.
This design separates three sources of uncertainty: whether the relevant moment
was localized, whether student behavior was recoverable, and whether that
behavior supports an ICAP label. Downstream agents can therefore abstain on a
missing modality while still using the remaining evidence. MAVEBench is intended
for segment-level feedback evaluation.

The benchmark deliberately preserves missingness. Camera-off tiles, occlusion,
unreadable slides, and ambiguous speakers are recorded as unavailable evidence,
not converted into Passive labels. Class-level proportions use only recoverable
student windows and are accompanied by the recoverable count. This prevents a
method from appearing confident simply because a modality failed. Because the
videos are public but contain identifiable classroom behavior, released artifacts
use anonymized identifiers and derived evidence rather than presenting identity
as an analytic variable.

\begin{figure*}[t]
\centering
\includegraphics[width=1.0\textwidth]{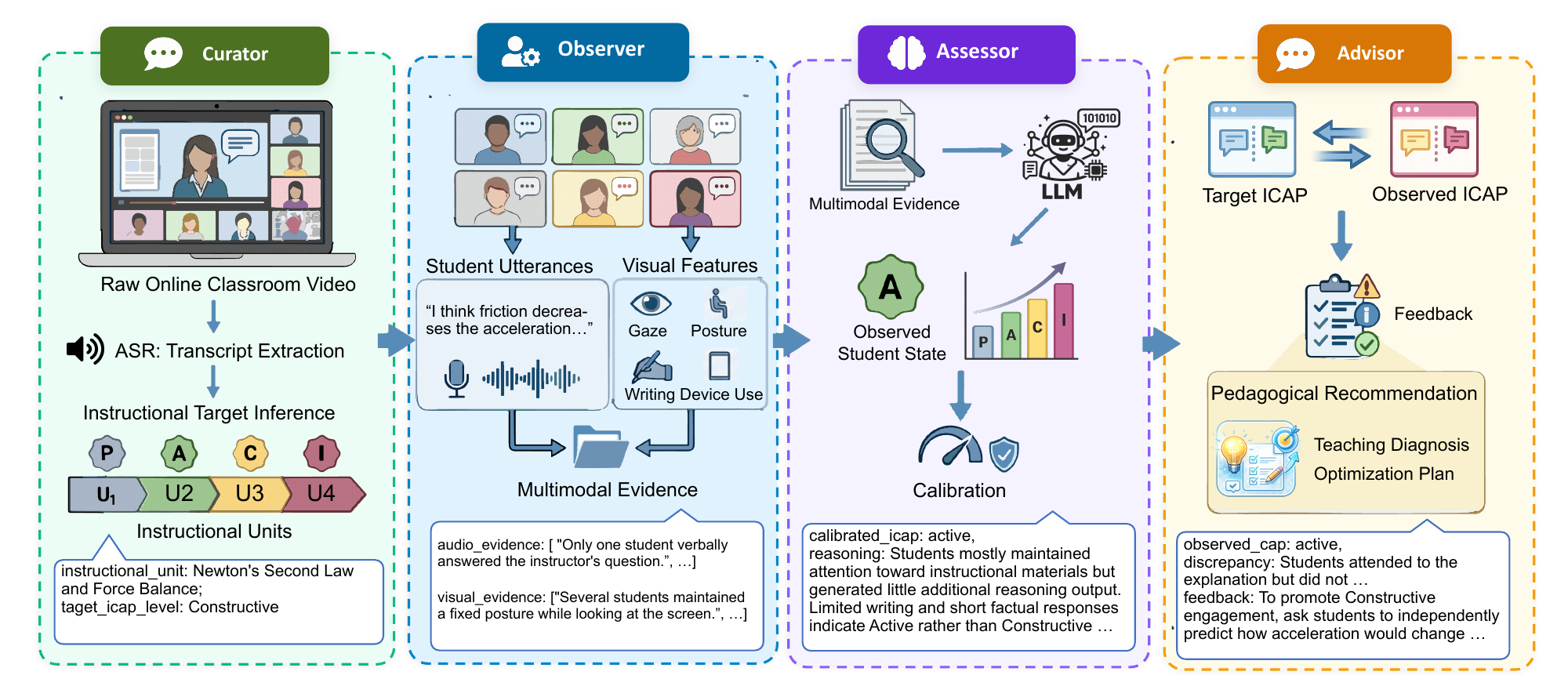}
\caption{Overview of MAVE. Given a raw online classroom video, the Curator segments the lecture and infers the intended ICAP objective for each instructional unit. The Observer extracts multimodal student evidence from the corresponding temporal window. The Assessor routes each student to an observed ICAP state and optionally calibrates a finer engagement level for Passive students in detail mode. Finally, the Advisor compares the intended target with the observed state distribution and generates time-aligned pedagogical feedback grounded in classroom evidence.} 
\label{fig:pipeline}
\end{figure*}

\subsection{Evaluation Protocol}
\label{sec:benchmark_evaluation}

Unlike conventional engagement analysis, pedagogical feedback is not valued only for predicting a student state. A useful recommendation must correctly diagnose the intended--observed gap, remain aligned with the instructional objective, and translate the diagnosis into an executable teaching move. We therefore evaluate MAVE with two complementary protocols: expert rubric
evaluation and controlled student-agent simulation.

\paragraph{Expert Rubric Evaluation.} The primary evaluation asks expert raters to score one evidence-grounded recommendation for each KP. All method outputs are anonymized and presented in a shared format. Experts rate each output on three 1--5 dimensions: (i) \emph{ICAP Accuracy}---whether the feedback identifies the observed cognitive-engagement state and the intended--observed gap; (ii) \emph{Alignment}---whether the proposed intervention preserves the KP objective and targets the requested transition; and (iii) \emph{Actionability}---whether the feedback provides concrete teacher language, timing, or steps that can be executed within the segment. Generic advice such as ''increase engagement'' cannot receive a high Actionability score unless it specifies both the activity and the observable evidence criterion. As complementary state-estimation metrics, we also compare the Assessor label with the expert label using classification accuracy and ordinal mean absolute error (MAE), mapping $P,A,C,I$ to $0,1,2,3$: \begin{equation} \mathrm{Acc}_{\mathrm{ICAP}}=\frac{1}{N}\sum_{k=1}^{N}\mathbb{1}[y_k=\hat y_k], \quad \mathrm{MAE}_{\mathrm{ICAP}}=\frac{1}{N}\sum_{k=1}^{N}|y_k-\hat y_k|. \end{equation} Accuracy treats all mistakes equally, whereas MAE distinguishes a neighboring mistake from a P--I reversal.

\par\vspace{2pt}\noindent\phantomsection
\label{sec:benchmark_simulation}%
\paragraph{Controlled Student-Agent Simulation.}
Expert rubrics assess whether feedback is well grounded and actionable, but they
do not directly test whether a recommendation can move students toward the
intended cognitive state. To evaluate this mechanism, we introduce a paired
student-agent simulation that measures state improvement relative to a controlled
baseline.

For each knowledge point $k$, let $t_k \in \{C,I\}$ denote the intended target
state. We consider two teacher conditions: a baseline teacher $\alpha$, which
receives the original excerpt and broad whole-class checks, and a guided teacher
$\beta$, which additionally receives MAVE's rewrite, scaffold, and focus
question. Each episode contains a hidden Session Controller, one Teacher agent,
and four Student agents initialized at $P,A,C,I$. The controller makes four
instructional decisions from \texttt{teaching}, \texttt{next\_page}, and
\texttt{interaction}. A student simulator $\phi$ produces student utterances
under teacher condition $m \in \{\alpha,\beta\}$:
\begin{equation}
U_k^m = \phi(k, T^m, s_0),
\end{equation}
where $T^m$ is the teacher input under condition $m$ and $s_0$ denotes the
shared initial student states. An independent judge $J$ labels the resulting
utterances using a withheld reference answer:
\begin{equation}
\hat{s}_k^m = J(U_k^m), \quad \hat{s}_k^m \in \{P,A,C,I\}.
\end{equation}

Absolute post-interaction state is not sufficient, because some students may
already begin near the target state. We therefore evaluate the contribution of
MAVE by comparing the guided condition against the paired baseline with
identical initial states:
\begin{equation}
\Delta_k = g(\hat{s}_k^\beta, t_k) - g(\hat{s}_k^\alpha, t_k),
\end{equation}
where $g(\hat{s},t)$ is a transition score, such as whether the
student reaches the target, the ordinal upgrade in cognitive level, or whether
the labeled utterance is at or above the target. Higher $\Delta_k$ indicates
that MAVE's recommendation induces stronger movement toward the intended state than the baseline teacher prompt.

\begin{table*}[htbp]
\centering
\caption{Expert human evaluation across frameworks (1--5 rubric).
``ICAP Accuracy'' is the expert-rated correctness of the identified ICAP
state. $p$: Welch's $t$-test of each baseline vs.\ MAVE on the Average score.}
\label{tab:main_results}

\resizebox{0.90\textwidth}{!}{%
\begin{tabular}{llccccc}
\toprule
\textbf{Framework} & \textbf{Domain} & \textbf{ICAP Accuracy} & \textbf{Actionability} & \textbf{Alignment} & \textbf{Average} & \textbf{$p$ (Avg.\ vs.\ MAVE)} \\
\midrule
Prompting & STEM & 3.50 $\pm$ 0.29 & 3.88 $\pm$ 0.70 & 3.50 $\pm$ 0.29 & 3.62 $\pm$ 0.36 & $<\!0.001$ \\
Reflexion & STEM & 3.36 $\pm$ 0.25 & 2.36 $\pm$ 0.25 & 3.55 $\pm$ 0.27 & 3.09 $\pm$ 0.22 & $<\!0.001$ \\
TextGrad & STEM & 3.91 $\pm$ 0.09 & 4.27 $\pm$ 0.82 & 4.00 $\pm$ 0.00 & 4.06 $\pm$ 0.15 & 0.001 \\
\textbf{MAVE (Ours)} & STEM & \textbf{4.82 $\pm$ 0.40} & \textbf{4.73 $\pm$ 0.65} & \textbf{4.45 $\pm$ 0.32} & \textbf{4.67 $\pm$ 0.33} & -- \\
\midrule
Prompting & Humanities & 3.50 $\pm$ 0.33 & 3.75 $\pm$ 0.25 & 3.50 $\pm$ 0.33 & 3.58 $\pm$ 0.25 & $<\!0.001$ \\
Reflexion & Humanities & 3.57 $\pm$ 0.53 & 2.57 $\pm$ 0.53 & 3.57 $\pm$ 0.53 & 3.24 $\pm$ 0.53 & $<\!0.001$ \\
TextGrad & Humanities & 4.00 $\pm$ 0.00 & 4.50 $\pm$ 0.28 & 4.10 $\pm$ 0.10 & 4.20 $\pm$ 0.03 & 0.008 \\
\textbf{MAVE (Ours)} & Humanities & \textbf{4.70 $\pm$ 0.48} & \textbf{4.80 $\pm$ 0.42} & \textbf{4.50 $\pm$ 0.53} & \textbf{4.67 $\pm$ 0.40} & -- \\
\bottomrule
\end{tabular}%
}

\end{table*}

\section{Method}

\paragraph{Overview.}
As shown in figure~\ref{fig:pipeline}, we formulate automated teaching feedback as a discrepancy reasoning problem between the instructor's intended ICAP objective and the students' observed ICAP state. Given a raw online classroom video $\mathcal{V}$, MAVE decomposes the problem into four coordinated agents: the \textbf{Curator} (pedagogical structuring and target inference), the \textbf{Observer} (multimodal evidence extraction), the \textbf{Assessor} (ICAP state estimation), and the \textbf{Advisor} (feedback generation), detailed in the subsections below.

\paragraph{Task Formulation.}
Formally, let the classroom video be partitioned into $K$ instructional segments
$\{\mathcal{V}_k\}_{k=1}^{K}$ and let $m\in\{1,\ldots,M_k\}$ index the
recoverable student windows in segment $k$. For each segment, MAVE estimates
(1) an intended ICAP target $c_k\in\mathcal{Y}$, where
$\mathcal{Y}=\{P,A,C,I\}$, (2) per-student observed states
$q_{k,m}\in\mathcal{Y}$, (3) optional Passive-student engagement levels
$e_{k,m}\in\{0,1,2,3\}$ in detail mode, and (4) a feedback text
$\mathcal{F}_k$ that explains the discrepancy between $c_k$ and the observed
classroom distribution.

The \textbf{Assessor} is responsible only for measuring student cognitive engagement. It first routes each student--segment pair through complementary linguistic and visual classifiers, and it invokes the finer engagement analysis only for students ultimately assigned to the Passive state in detail mode. The \textbf{Advisor} is responsible only for acting on the pedagogical gap: given the intended level, the distribution of observed ICAP states, and the passive-student engagement summary, it synthesizes feedback. This separation keeps diagnostic measurement independent from recommendation generation.

\subsection{Curator: Pedagogical Structuring}
The Curator reconstructs the pedagogical structure of the lecture from the raw classroom stream, which aims to infer the cognitively appropriate ICAP objective. In other words, the agent estimates what level of student cognitive engagement a particular knowledge unit should ideally elicit according to pedagogical principles.

\paragraph{Instructional parsing.}
As an auxiliary tool, the Curator first invokes a Transformer-based automatic speech recognition (ASR) module to obtain a timestamp-aligned transcript $ \mathcal{T} = \{(w_n, t_n)\}_{n=1}^{N} $, where $w_n$ denotes the recognized token and $t_n$ denotes its temporal boundary. The transcript is aligned with the classroom timeline and grouped into contiguous instructional windows.

\paragraph{Instructional unit construction.}
Based on the transcript and temporal structure, the Curator partitions the lecture into pedagogically coherent instructional units $\mathcal{U}_k = (\mathcal{W}_k, \tau_k)$, where $\mathcal{W}_k$ denotes the instructional content within the segment and
$\tau_k = [t_k^{s}, t_k^{e}]$
denotes its temporal interval.
Each instructional unit corresponds to a relatively self-contained knowledge delivery episode. This segmentation establishes a stable temporal abstraction for downstream multimodal reasoning and enables direct alignment between instructional context and student responses.

\par\vspace{2pt}\noindent\textit{ICAP target inference.}
For each instructional unit $\mathcal{U}_k$, the Curator analyzes the instructional content, discourse structure, and pedagogical interaction patterns to infer the cognitively appropriate ICAP target $c_k \in \{P,A,C,I\}$. Instead of predicting what the teacher subjectively intended, the Curator estimates what form of cognitive engagement the knowledge component should ideally promote. For example, factual exposition may correspond to Passive or Active engagement, while conceptual synthesis, reasoning prompts, or collaborative problem solving are expected to induce Constructive or Interactive engagement. The inferred ICAP target serves as the pedagogical reference state for the subsequent discrepancy analysis.

In practice, feedback should specify a realizable next step rather than demand
maximal interaction everywhere. We therefore use Constructive (C) and Interactive (I)
states as actionable targets: Constructive engagement asks each learner to
generate an explanation, prediction, or solution beyond the presented material,
whereas Interactive engagement additionally requires responsive knowledge
construction across learners. The target combines conceptual demand, lesson
position, and evidence of readiness. Concretely, the rule-based gate forces an
Interactive target when Interactive behavior is already present or the
recoverable Constructive coverage is at least 50\%; otherwise it forces a
Constructive target. A soft early-stage safeguard overrides Interactive for the
first 30\% of a lecture, when exposition and orientation are still dominant. The
Curator then supplies the target-specific rationale and question, so an
Interactive target cannot degenerate into a generic recommendation to ``add
discussion.''

\refstepcounter{subsection}
\par\vspace{2pt}\noindent{\large\bfseries
\thesubsection\quad Observer: Learner Feedback Modeling}\par\nobreak\vspace{2pt}

Rather than predicting engagement labels directly, the Observer converts
multimodal classroom signals into temporally aligned natural-language evidence
for downstream pedagogical reasoning.

\paragraph{Audio interaction parsing.}
For each instructional unit $\mathcal{U}_k$, the Observer invokes an automatic speech recognition (ASR) module to obtain timestamp-aligned classroom transcripts. Based on speaker turns and interaction context, the Observer further separates student utterances from the instructor's speech, yielding a student-side interaction transcript $\mathcal{A}_k = \{(s_i, t_i)\}$,
where $s_i$ denotes a student utterance and $t_i$ denotes its temporal location. These parsed responses provide linguistic evidence about participation, reasoning behaviors, and collaborative interaction patterns.

\paragraph{Visual evidence extraction.}
For the visual stream, the Observer invokes a video parsing tool to decompose the online classroom layout into individual student sub-windows. Frames are then sampled from each student region at fixed temporal intervals within $\tau_k$. The sampled frames are converted into factual natural-language descriptions that summarize observable behaviors such as gaze direction, posture, writing activity, device usage, and interaction with learning materials.

Formally, the extracted visual evidence for the $k$-th instructional unit is represented as $\mathcal{B}_k = \{v_k^{1}, v_k^{2}, \dots, v_k^{M_k}\}$,
where each $v_k^{m}$ denotes the textual behavioral description of the $m$-th student window.

The Observer aggregates the linguistic and visual evidence into a temporally aligned multimodal evidence summary $\mathcal{E}_k = \{\mathcal{A}_k, \mathcal{B}_k\}$.
This representation preserves local classroom behaviors while reducing frame-level noise, thereby providing a stable foundation for downstream ICAP assessment and discrepancy reasoning.

Evidence is aligned by instructional segment and student window rather than
pooled across the recording. Delayed answers, note taking, and gaze changes
retain timestamps so that a response remains associated with the prompt that
elicited it. Missing cameras, occluded faces, presentation-only frames, and
ambiguous speakers are marked unavailable or uncertain instead of becoming
negative evidence. The Observer describes what is visible or audible but does
not infer ICAP; this provenance boundary makes later decisions auditable and
limits behavioral over-interpretation.

\subsection{Assessor: ICAP State Estimation}

The Assessor estimates an ICAP state for each student and instructional unit
rather than assigning one label to the whole classroom. Its outputs are consumed
by the Advisor and are not used to generate recommendations directly. We use a
conservative routing rule with three complementary prompts. A discourse prompt
returns ``Passive/Active'' when no student speech is detected, ``Constructive''
when one student produces constructive output, and ``Interactive'' only when at
least two students contribute constructive turns with evidence of turn taking. A
separate visual-action detector resolves the ``Passive/Active'' case to Active
when overt learning-related motor activity (e.g., writing, drawing, or
manipulating a learning object) is observed; otherwise the student remains
Passive. Thus, a silent student is not promoted to Constructive or Interactive
merely because a face is visible or a posture is stable.

\paragraph{Detail-mode Passive monitoring.}
When the routed state is $P$ and detail mode is enabled, the Observer's cleaned
student crop is analyzed at a finer temporal resolution. Frames are sampled at
fixed intervals, aggregated into one-minute factual narratives, and grouped into
non-overlapping three-minute windows. For student $m$ in window $t$, an
engagement judge selects an ordinal anchor $a_{m,t}\in\{0,1,2,3\}$ based on
observable evidence and makes binary plausibility judgments for neighboring
levels. Specifically, $a_{m,t}=0$ indicates frequent looking away/down or
obvious inattentiveness; $a_{m,t}=1$ indicates weak or unstable attention,
such as inconsistent orientation or posture; $a_{m,t}=2$ indicates generally
forward and stable attention with only minor lapses; and $a_{m,t}=3$ indicates
stronger observable engagement cues, such as nodding, leaning in, speaking, or
active hand movements without sustained distraction. These levels refine
engagement \emph{within} the Passive state and do not alter the student's
ICAP classification.
\paragraph{Ordinal calibration.}
To reduce local LLM variability, the binary judgments are converted into an ordinal-consistent distribution. Let $d_{m,t}(j)$ be zero for the anchor, one (two) for an acceptable one-level (two-level) deviation, and a maximal penalty otherwise, with $d_{m,t}(j)\geq |j-a_{m,t}|$. We use
\begin{equation}
w_{m,t}(j)=\exp\!\left[-\alpha d_{m,t}(j)\right], \qquad
p_{m,t}(j)=\frac{w_{m,t}(j)}{\sum_{r=0}^{3}w_{m,t}(r)},
\end{equation}
and select the risk-minimizing engagement level
\begin{equation}
e_{m,t}=\arg\min_{j\in\{0,1,2,3\}}\sum_{r=0}^{3}|j-r|p_{m,t}(r).
\end{equation}
In our implementation $\alpha=0.7$. A candidate deviation receives distance
one or two only when the judge accepts the corresponding one- or two-level
alternative; rejected deviations receive a maximal penalty of three. We then
enforce $d_{m,t}(j)\geq |j-a_{m,t}|$ before normalization. Windows with no
usable student evidence (e.g., presentation-only or excluded frames) are
skipped rather than assigned an engagement score. The Advisor receives both
the ICAP-state proportions and the distribution of $e_{m,t}$ among Passive
students, allowing it to distinguish a broad Passive bottleneck from a small
number of highly distracted individuals.

The detail branch follows a fixed evidence path. The cleaner first removes
black, icon-only, and presentation-dominated frames. The visual narrator then
summarizes up to six sampled frames from each one-minute interval in factual
prose, and consecutive summaries are grouped into non-overlapping three-minute
windows. The judge reads the three summaries, selects one anchor level, and
returns binary plausibility judgments for neighboring ordinal levels. A
two-level deviation cannot be accepted when its corresponding one-level
deviation is rejected. The prompt instructs the judge to use only visible
behavior, select one 0--3 anchor from the overall three-minute tendency, avoid
defaulting to level 2, and return the anchor, four binary deviation judgments,
and a brief evidence-grounded reason as strict JSON. Stable forward orientation
or strong focus cues support level 3, whereas recurring distraction supports
levels 0 or 1.

Aggregating per-student decisions also exposes \emph{coverage}: a segment may
contain one strong constructive answer while most of the class remains Passive.
MAVE reports both the highest supported ICAP behavior and the fraction of
recoverable students at each state, avoiding the common error of treating a
vocal minority as evidence that the objective was achieved class-wide.

\begin{figure*}[t]
\centering
\begin{subfigure}[b]{0.33\textwidth}
  \centering
  \includegraphics[width=\linewidth]{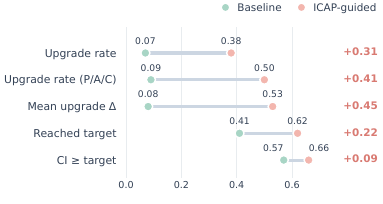}
  \caption{Guided ICAP transitions.}
  \label{fig:icap_simulation}
\end{subfigure}\hfill
\begin{subfigure}[b]{0.33\textwidth}
  \centering
  \includegraphics[width=\linewidth]{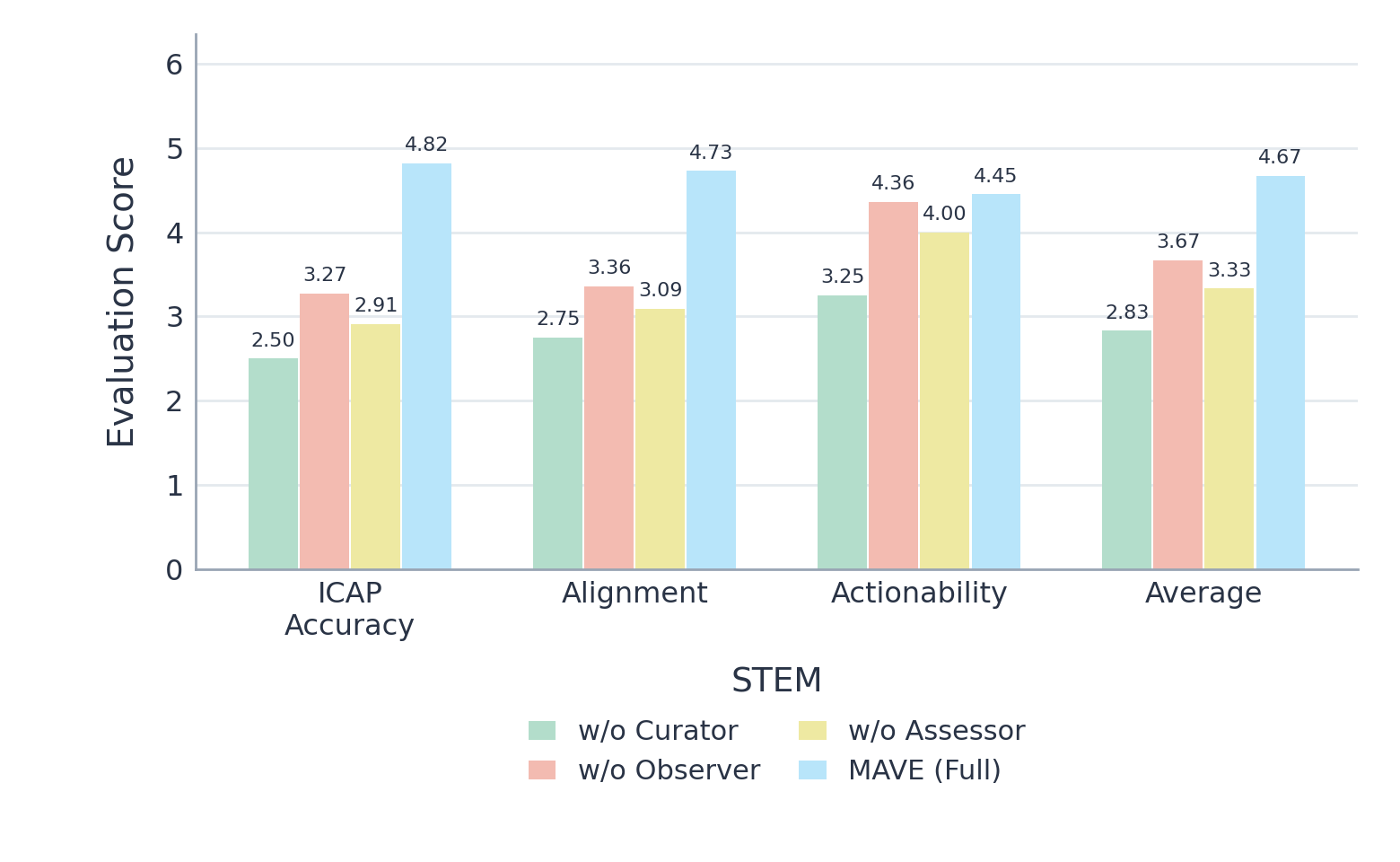}
  \caption{Ablation (STEM).}
  \label{fig:ablation_stem}
\end{subfigure}\hfill
\begin{subfigure}[b]{0.33\textwidth}
  \centering
  \includegraphics[width=\linewidth]{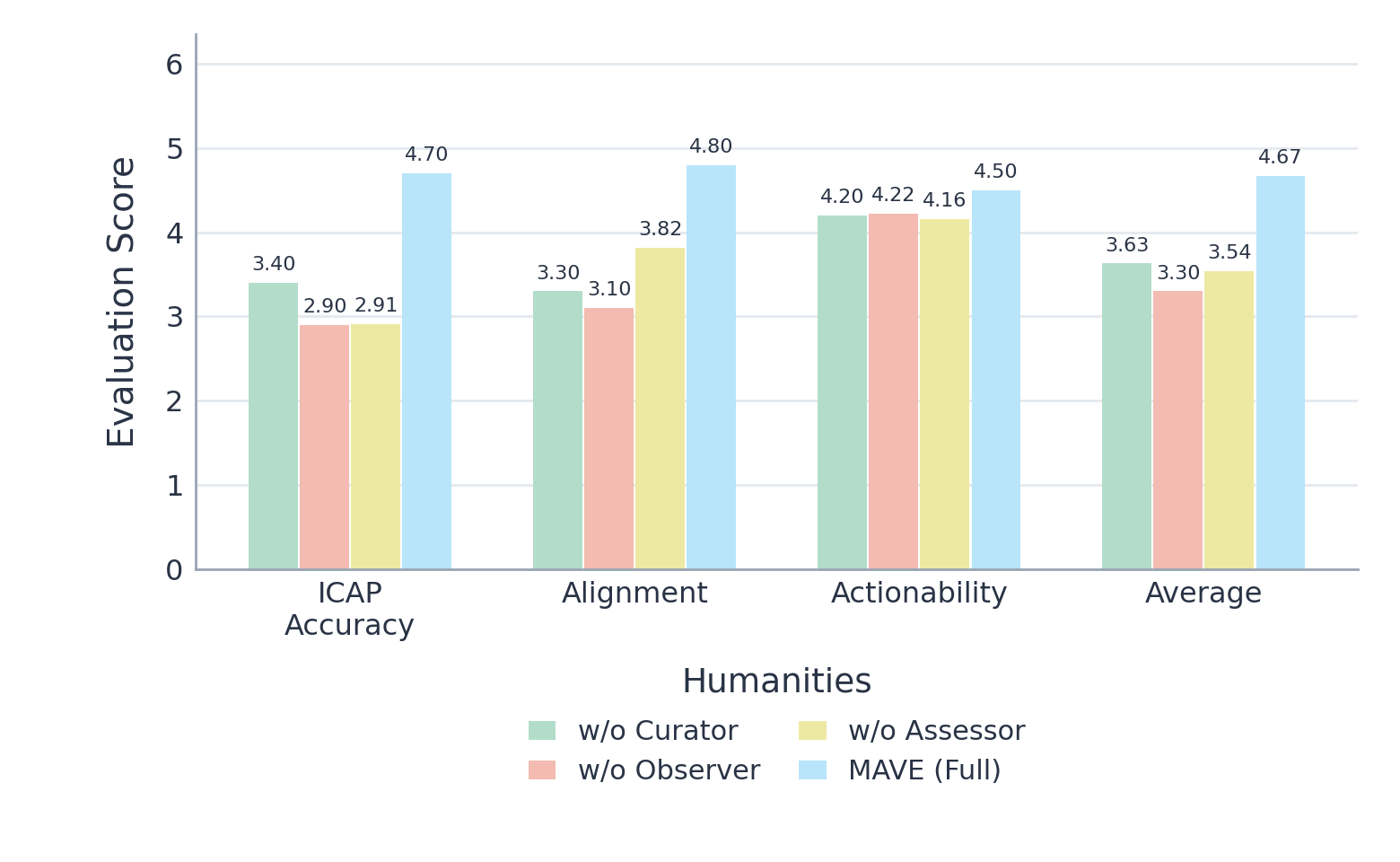}
  \caption{Ablation (Humanities).}
  \label{fig:ablation_hum}
\end{subfigure}
\vspace{-0.4em}
\caption{\textbf{Additional analyses.} (a)~Student-Agent Transition simulation: per-metric means, Baseline vs.\ ICAP-guided,
with paired deltas. (b,~c)~Ablations: removing the Curator, Observer, or
Assessor degrades all expert-rated metrics vs.\ full MAVE on both domains.}
\vspace{-0.8em}
\label{fig:main_additional_results}
\end{figure*}

\subsection{Advisor: Feedback Generation}

The Advisor synthesizes teacher-side context, Assessor outputs, and student evidence into time-aligned teaching recommendations. It does not re-estimate ICAP levels from raw video; it reasons over $c_k$, the empirical distribution of $q_{k,m}$, the optional Passive-student engagement levels $e_{k,m}$, and $\mathcal{E}_k$ to explain why a segment succeeded or failed relative to the intended pedagogical objective.

\paragraph{Pedagogical Gap Diagnosis.}
For each instructional unit $\mathcal{U}_k$, the Advisor compares the intended
ICAP target $c_k$ with the observed state distribution
$\{q_{k,m}\}_{m=1}^{M_k}$. The resulting gap indicates whether the instructional
design and the actual student response are aligned:
\begin{equation}
\Delta_k = \phi\!\left(c_k,\{q_{k,m}\}_{m=1}^{M_k},
\{e_{k,m}:q_{k,m}=P\}\right),
\end{equation}
where $\phi(\cdot)$ measures the pedagogical gap between intended and observed
cognitive states, including coverage of the target state and, when available,
the within-Passive engagement profile.

The gap is directional and coverage-aware. If the target is Constructive but
the class is predominantly Passive, the first priority is to make thinking
externally observable with an individual prediction or explanation; peer
interaction is not yet required. If several learners reach Constructive but
coverage remains low, the Advisor expands participation before raising the
cognitive level. When Constructive engagement is broadly established and the
target is Interactive, the recommendation instead introduces turn-taking in
which students build on, challenge, or reconcile peers' contributions.
Conversely, if the target has already been reached, the Advisor preserves the
successful activity and suggests only a small refinement. This ordering
distinguishes increasing \emph{coverage} at the current level from moving a
smaller set of learners to a higher one.

\paragraph{Recommendation synthesis.}
Conditioned on the instructional content $\mathcal{U}_k$, the student evidence $\mathcal{E}_k$, and the discrepancy $\Delta_k$, the Advisor generates a time-local recommendation:
\begin{equation}
\mathcal{F}_k = \Psi(\mathcal{U}_k, \mathcal{E}_k, \Delta_k).
\end{equation}
The generated feedback is grounded in concrete classroom evidence and is explicitly aligned with the intended ICAP transition. For example, when the instructor intends to elicit Constructive or Interactive engagement but the observed state remains Passive or Active, the Advisor can recommend adding prompts, peer discussion, wait time, or more scaffolded questioning.

Each output contains a diagnosis, a concise rewrite of the relevant teaching
move, one or two operational scaffolds, a target-level question, and a reference
answer. These fields force the recommendation to identify \emph{where} the gap
occurred, \emph{what} the instructor can say or do, and \emph{what evidence}
would demonstrate improvement. For a Constructive target, the proposed question must elicit
new learner-generated content without relying on peer exchange. For an Interactive target,
it must require at least two learners to respond to one another. The reference
answer constrains domain correctness, while citations to the transcript and
student evidence constrain behavioral grounding.

\begin{table}[t]
\centering
\caption{Consistency between expert human and LLM-as-a-Judge evaluation on
the double-annotated reliability subset. Agreement is normalized score
proximity; $p$: Welch's $t$-test, human vs.\ LLM.}
\label{tab:judge_consistency}
\setlength{\tabcolsep}{2.5pt}
\tablebodyfont
\renewcommand{\arraystretch}{0.92}
\begin{tabular}{llcccc}
\toprule
Metric & Domain & LLM Score & Human Score & Agreement & $p$ \\
\midrule
\multirow{2}{*}{ICAP Accuracy} & STEM & 4.95 $\pm$ 0.04 & 4.82 $\pm$ 0.40 & 90.9\% & 0.31 \\
& Humanities & 4.90 $\pm$ 0.09 & 4.70 $\pm$ 0.48 & 80.0\% & 0.23 \\
\midrule
\multirow{2}{*}{Actionability} & STEM & 5.00 $\pm$ 0.00 & 4.73 $\pm$ 0.65 & 81.8\% & 0.37 \\
& Humanities & 4.80 $\pm$ 0.16 & 4.80 $\pm$ 0.42 & 85.0\% & 1.00 \\
\midrule
\multirow{2}{*}{Alignment} & STEM & 4.22 $\pm$ 0.28 & 4.45 $\pm$ 0.52 & 81.8\% & 0.22 \\
& Humanities & 4.60 $\pm$ 0.24 & 4.50 $\pm$ 0.53 & 85.3\% & 0.60 \\
\bottomrule
\end{tabular}
\end{table}

\section{Experiment}

We conduct comprehensive experiments to validate MAVE's ability to generate
evidence-grounded and actionable pedagogical feedback from online classroom
videos.

Our evaluation addresses four key research questions: \textbf{(RQ1)} Does MAVE improve pedagogical feedback quality
    over baselines? \textbf{(RQ2)} Which specialized agents contribute to MAVE's
    improvement? \textbf{(RQ3)} Can MAVE-guided recommendations induce the intended
    cognitive-state transitions in a controlled student-agent simulation? \textbf{(RQ4)} How sensitive are MAVE's quality and cost to different
    backbone models?

\subsection{Experiment Setup}
We evaluate all methods on MAVEBench, which contains 173 KP-level instructional
segments from approximately 20 hours of authentic online-class videos 
(Section~\ref{sec:dataset}). Following the evaluation protocol in
Section~\ref{sec:benchmark_evaluation}, we report two complementary evaluations:
offline expert rubric evaluation for feedback quality and controlled
student-agent simulation for state transitions. For offline evaluation, expert scores are computed on the 173 rated KP-level
outputs by five expert evaluators with university-level teaching experience. All methods receive the same anonymized evidence, KP boundaries,
transcripts, output schemas, decoding settings, retry policy, and GPT-4o-mini
backbone. This matched setup ensures that differences reflect agent
specialization, routing, and gap-conditioned reasoning rather than differences
in preprocessing or perceptual inputs. The controlled simulation uses the same
173 KPs with paired baseline/guided conditions and five random seeds; only this
simulation uses GPT-5 for student-agent runs.

\subsection{Comparisons and Baselines}
\label{sec:comparisons_baselines}

Because evidence-grounded pedagogical feedback from online-class videos is a
new task, there is no established education-specific baseline that directly
matches our input-output setting. We therefore adapt several general feedback
generation and refinement methods to this domain, while keeping the evidence,
instructional context, and output schema identical across methods. We compare with a directly prompted VLM (\textit{Prompting}),
Reflexion~\cite{shinn2023reflexion}, and
TextGrad~\cite{yuksekgonul2024textgrad}. Each receives the same transcript,
instructional context, student evidence, and output schema. Prompting tests
direct evidence-to-advice generation; Reflexion adds iterative self-critique
and memory; TextGrad optimizes a draft with textual feedback. The comparison
thus spans one-pass, reflective, and optimization-based generation. Only MAVE
explicitly separates target inference, evidence description, state estimation,
and gap-conditioned advising.

\subsection{Main Results (RQ1)}
\label{sec:main_results}

Table~\ref{tab:main_results} reports expert-rated feedback quality, showing that MAVE leads all baselines on all three metrics in both domains. Its average is 4.67 versus 4.06 for the strongest STEM baseline and 4.67 versus 4.20 in Humanities, gains of 0.58 and 0.43 points. The largest consistent separation is in ICAP Accuracy: MAVE reaches 4.82/4.70, whereas the strongest baseline reaches 3.91/4.00. TextGrad narrows the Actionability gap, showing that iterative optimization can produce well-formed interventions, but its weaker diagnosis limits their grounding. Reflexion's lower Actionability likewise shows that more critique does not guarantee executable advice. MAVE's nearly identical STEM and Humanities averages (4.67 vs. 4.67) indicate that its structured gap representation transfers across subject matter. Together with Fig.~\ref{fig:main_additional_results}, the results attribute the gain to explicit targets and student evidence rather than fluency alone.

\paragraph{Evaluation Reliability}
\label{sec:metrics_evaluation}
Section~\ref{sec:benchmark_evaluation} specifies the evaluation method and
metrics. We report expert judgments as the
primary evidence and use the independent LLM judge only to check whether the
same rubric can be applied consistently at scale. Average scores are computed
from unrounded per-item values. Table~\ref{tab:judge_consistency} uses the double-annotated reliability subset,
so its human means differ slightly from the full-set means in
Table~\ref{tab:main_results}. Human and LLM means do not differ significantly
on any metric in either domain (Welch's $t$ on per-item scores, all $p>0.2$).
``Agreement'' is normalized score proximity. For human and LLM scores $h_i$
and $\ell_i$ on the double-annotated subset, it is
$100[1-N^{-1}\sum_i|h_i-\ell_i|/5]$.


\begin{table}[t]
\centering
\caption{Student-Agent Transition Simulation: paired per-KP results over 173 KPs and
five seeds.}
\label{tab:transition_results}
\setlength{\tabcolsep}{2.5pt}
\tablebodyfont
\renewcommand{\arraystretch}{0.92}
\begin{tabular}{lcccccc}
\toprule
Metric & Baseline & Guided & $\Delta$ & 95\% CI & $d$ & $p$ \\
\midrule
Upgrade Rate & 0.069 & 0.375 & +0.307 & [+0.289,+0.324] & 2.63 & $<10^{-3}$ \\
Upgrade Rate (P/A/C) & 0.091 & 0.501 & +0.409 & [+0.386,+0.433] & 2.63 & $<10^{-3}$ \\
Mean Upgrade & 0.078 & 0.525 & +0.447 & [+0.420,+0.475] & 2.44 & $<10^{-3}$ \\
Reached Target & 0.406 & 0.623 & +0.217 & [+0.196,+0.239] & 1.53 & $<10^{-3}$ \\
CI $\geq$ Target & 0.566 & 0.659 & +0.093 & [+0.054,+0.132] & 0.36 & $<10^{-3}$ \\
\bottomrule
\end{tabular}
\end{table}

\subsection{Component Analysis (RQ2)}
Figure~\ref{fig:main_additional_results}b--c removes the Curator, Observer, or
Assessor while holding the backbone, inputs, and rubric fixed. Every removal
reduces the average score in both domains (per-item Welch tests, all
$p<0.05$). Removing the Curator causes the
largest overall degradation because the advice no longer has an intended-target
reference. Removing the Observer most directly weakens ICAP Accuracy:
transcript-only reasoning can recognize a spoken answer but cannot determine how
broadly participation extends across the class. Removing the Assessor makes
results less stable because raw evidence reaches the Advisor without conservative
P/A/C/I routing or Passive calibration. Actionability is comparatively resilient,
showing that a model can write a plausible activity without correctly diagnosing
why it is needed. The full pipeline makes that activity evidence-grounded and
level-appropriate.

The magnitude and location of these drops separate the agents' roles. Relative
to full MAVE, removing the Curator lowers Average by 1.81 points in STEM and
1.00 in Humanities; removing the Observer lowers it by 0.97 and 1.33 points.
The larger Humanities penalty for the Observer is consistent with
discussion-centered segments, where distributed verbal and visible responses
are needed to estimate coverage. Removing the Assessor causes intermediate
losses of 1.31 and 1.09 points. Intended-level inference is therefore most
consequential in STEM, whereas multimodal coverage is especially consequential
in Humanities. Passing raw observations directly to the Advisor cannot replace
an explicit, calibrated state summary. This asymmetric pattern also cautions
against treating Average alone as a complete explanation: the same pipeline
stage can fail differently as instructional evidence changes across domains.

\begin{figure*}[t]
\centering

\begin{minipage}[c]{0.60\textwidth}
    \centering
    \captionof{table}{Efficiency per video over the 20 MAVEBench videos.
    Tokens combine input and output; latency is wall-clock time per model
    request, time is the end-to-end pipeline time, and cost uses the observed
    token mix.}
    \label{tab:efficiency_small}

    \setlength{\tabcolsep}{2.5pt}
    \tablebodyfont
    \renewcommand{\arraystretch}{0.92}

    \resizebox{\linewidth}{!}{%
    \begin{tabular}{lcccc}
        \toprule
        Method & Tokens (M) & Latency (s) & Time (min) & Cost (\$) \\
        \midrule
        Prompting & 0.02 $\pm$ 0.02 & 1.4 $\pm$ 0.2 & 0.03 $\pm$ 0.01 & 0.01 $\pm$ 0.01 \\
        Reflexion & 1.69 $\pm$ 0.04 & 1.8 $\pm$ 0.3 & 0.15 $\pm$ 0.04 & 0.76 $\pm$ 0.02 \\
        TextGrad & 1.20 $\pm$ 0.02 & 1.8 $\pm$ 0.3 & 0.12 $\pm$ 0.03 & 0.54 $\pm$ 0.01 \\
        MAVE & 0.75 $\pm$ 0.07 & 1.9 $\pm$ 0.4 & 0.20 $\pm$ 0.05 & 0.34 $\pm$ 0.03 \\
        MAVE detailed & 2.22 $\pm$ 0.54 & 2.5 $\pm$ 0.5 & 0.65 $\pm$ 0.18 & 1.00 $\pm$ 0.16 \\
        \bottomrule
    \end{tabular}%
    }
\end{minipage}
\hfill
\begin{minipage}[c]{0.36\textwidth}
    \centering
    \includegraphics[width=1.10\linewidth]{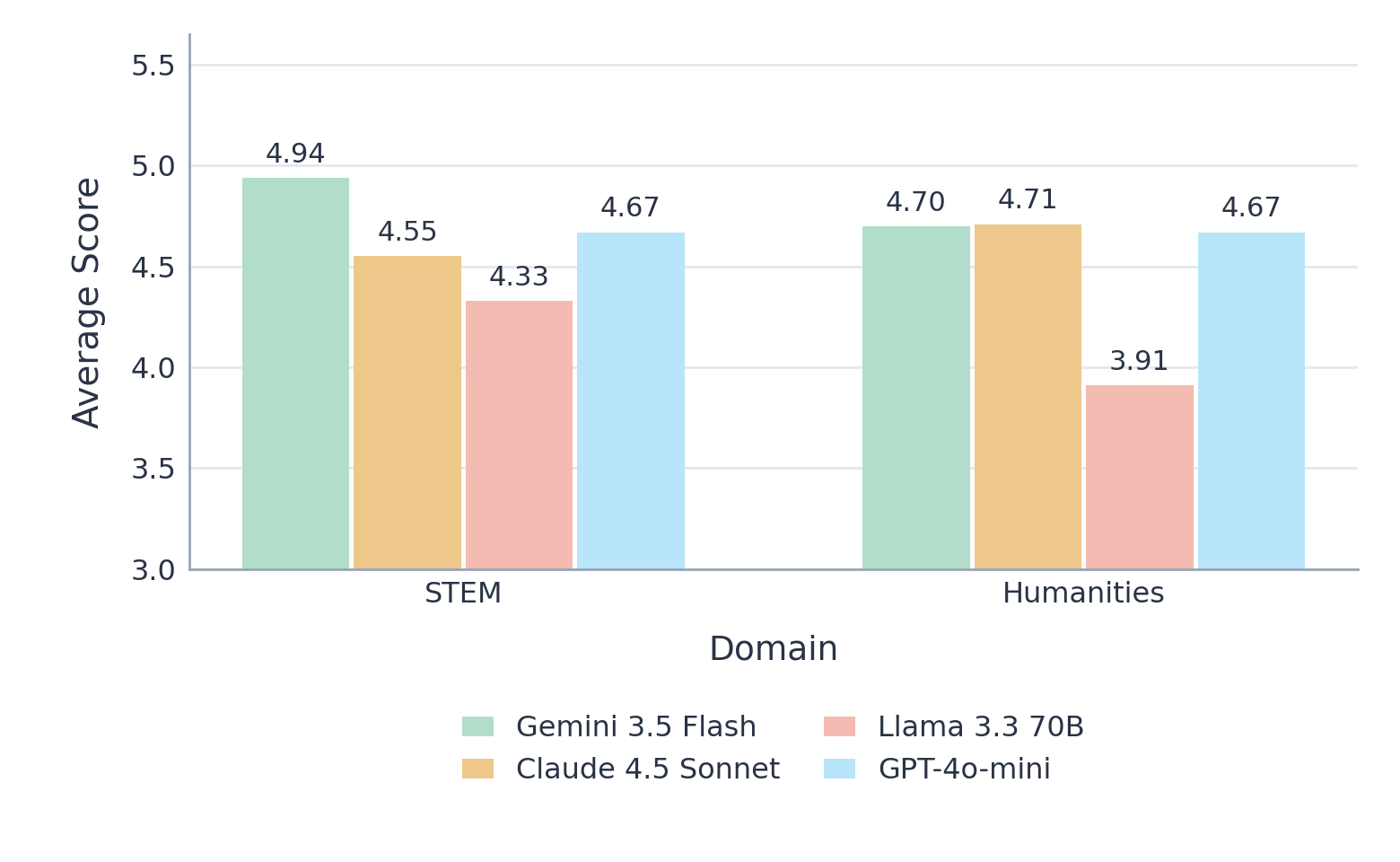}
    \captionof{figure}{Average performance across backbone LLMs and domains.}
    \label{fig:backbone_sensitivity}
\end{minipage}

\end{figure*}

\subsection{Student-Agent Transition Simulation (RQ3)}
\label{sec:icap_transition}
In a separate controlled simulation covering all 5.8K cleaned frames and 110K face crops. (the same 173 KPs;
five seeds; 865 paired KP--seed comparisons),
GPT-5-simulated students receiving ICAP-guided instruction upgrade more often
and further: Upgrade Rate rises from 0.069 to 0.375, Mean Upgrade from 0.078 to
0.525 ICAP levels, and the share of students reaching the target mode from
0.406 to 0.623, with large paired effect sizes (Cohen's $d$ up to 2.63;
Fig.~\ref{fig:icap_simulation}). These simulation results are reported
separately from the offline video scores. The paired setup and metric definitions appear in
Section~\ref{sec:benchmark_simulation}. Guidance improves
all five metrics (Table~\ref{tab:transition_results}), including upgrade rate
among P/A/C starters (0.091 to 0.501). The smaller gain in utterance-level
CI-at-target (+0.093) than in student-level upgrades indicates that MAVE mainly
broadens and deepens participation rather than merely increasing high-level
phrases. Because participants and judge are simulated, this tests the mechanism
under controlled conditions, not realized learning gains. Each pair starts from the same KP, states, seed, and reference answer. Stronger
student-level than utterance-level gains are consistent with first broadening
participation or moving a learner one level. Table~\ref{tab:transition_results} reports paired per-KP results. Every
difference is significant under paired
$t$-tests over the 173 KPs ($\mathrm{df}=172$; largest
$p=4.9{\times}10^{-6}$).

\subsection{Sensitivity Analysis (RQ4)}
Replacing the backbone with Gemini 3.5 Flash, Claude 4.5 Sonnet, or Llama
3.3 70B keeps performance strong and broadly stable across backbones. This
suggests MAVE's advantage is structural---multimodal grounding and discrepancy
reasoning---not an artifact of GPT-4o-mini.

Gemini and Claude remain close to the GPT-4o-mini reference, while Llama 3.3
70B declines most clearly, especially on Humanities Alignment
(Fig~\ref{fig:backbone_sensitivity}). MAVE therefore transfers across model
families, although weaker backbones still affect domain reasoning. Efficiency
shows a complementary trade-off (Table~\ref{tab:efficiency_small}): standard
MAVE averages 0.75M tokens and \$0.34 per video, below iterative Reflexion and
TextGrad in token use and cost but above direct Prompting. Detail mode rises to
2.22M tokens and \$1.00 because Passive cases require repeated frame narratives
and ordinal judgments. Since A/C/I cases bypass this branch, it can be reserved
for segments where diagnosing widespread Passive behavior warrants the cost.

Only the additional within-Passive trajectory is deferred in the conditional
detail branch. An instructor can request the cheaper segment-level diagnosis
first and expand selected cases when distinguishing inattention from quiet but
sustained participation changes the recommendation. Relative to GPT-4o-mini, only Llama~3.3~70B degrades significantly
(Humanities $p<0.001$, STEM $p=0.04$); Gemini is significantly higher only on
STEM ($p=0.04$), while Claude and the remaining comparisons are not
significant ($p\geq0.05$). Figure~\ref{fig:backbone_sensitivity} visualizes the
cross-backbone averages, and gives the
domain-specific scores.

\subsection{Efficiency}
\label{sec:efficiency}
Table~\ref{tab:efficiency_small} reports per-video means over MAVEBench, with
input and output tokens combined. Cost uses the pipeline's input/output mix
(blended approximately \$0.45/M tokens). Standard MAVE remains feasible for
offline analysis. Detailed MAVE adds frame-to-narrative calls and one ordinal
judgment per three-minute Passive window; A/C/I students do not incur this
refinement.

\begin{figure*}[!t]
    \centering
    \includegraphics[width=1.0\linewidth]{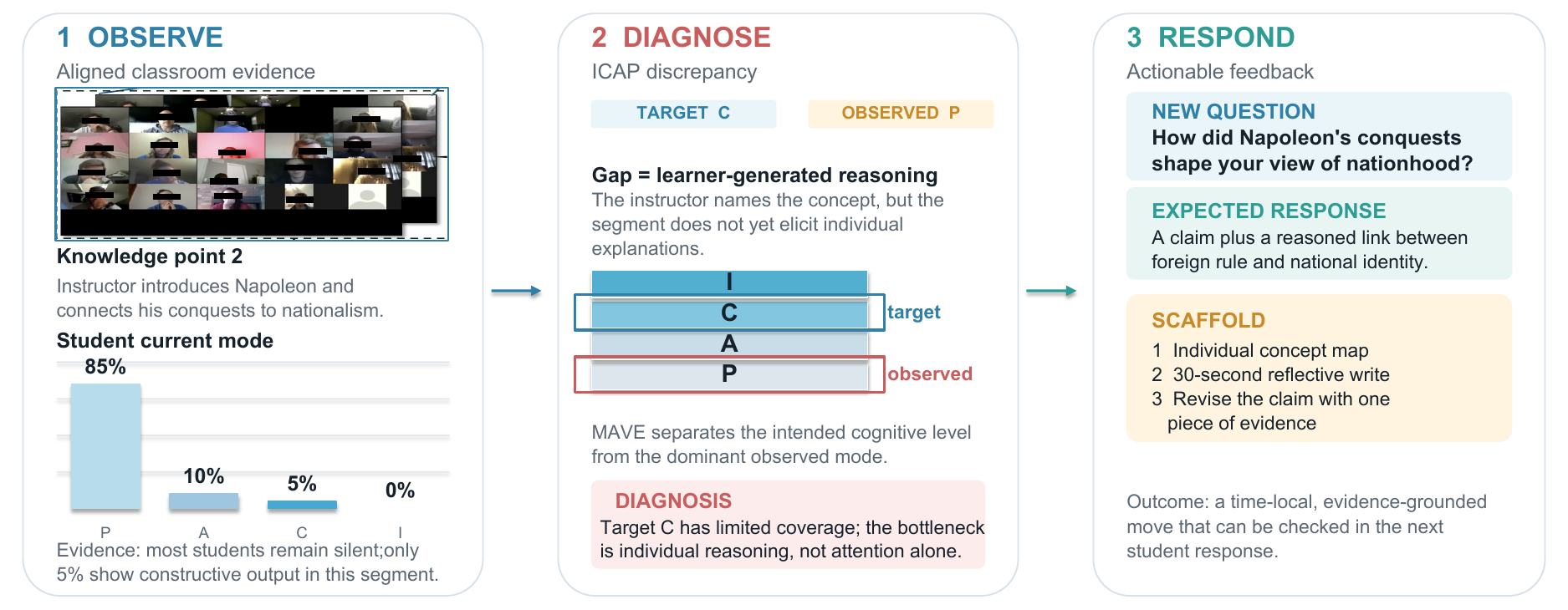}
    \caption{Illustrative MAVE case study. Multimodal observations support the
    inferred ICAP gap and the resulting time-local pedagogical recommendation.}
    \label{fig:case_study}
\end{figure*}

\subsection{Qualitative Case Study}
Figure~\ref{fig:case_study} illustrates a representative Humanities segment in which
the instructor explains a multi-step solution and the Curator assigns a
Constructive target. The Observer finds limited student speech and infrequent
note taking across the recoverable student windows; the Assessor consequently
routes the segment to a predominantly Passive state. Rather than responding
with the generic instruction to ``increase interaction,'' the Advisor localizes
the gap to the intermediate reasoning step, rewrites it as an individual
prediction-and-explanation prompt, supplies a scaffold, and states a reference
answer against which the response can be checked. This example shows how the
agent interfaces preserve a trace from visible or audible evidence, through an
ICAP diagnosis, to a teacher action.

Prompting offers a broad participation suggestion without tying it to the
observed behavior. Reflexion improves the wording but does not identify which
cognitive output is missing. TextGrad produces a more structured activity, yet
parts of it are weakly connected to the local evidence. MAVE's advantage is
therefore not simply longer feedback: its recommendation identifies the target,
the coverage bottleneck, and the observable response that would close the gap.

\section{Discussion and Conclusion}

We presented \textbf{MAVE}, a multi-agent framework that grounds pedagogical
feedback in ICAP theory and multimodal classroom evidence. The central idea is
to represent feedback as a gap between the cognitive activity appropriate for
an instructional segment and the activity supported by student evidence in that
same segment. This shifts classroom analytics from aggregate descriptions of
attention to a specific, actionable diagnosis of where instructional intent and
student response diverge. MAVE operationalizes this idea through inspectable intermediate artifacts. The
Curator constructs knowledge-point units and ICAP targets; the Observer extracts
time-aligned speech and visible behavior while preserving missingness; the
Assessor conservatively estimates ICAP states and, when needed, refines
Passive cases with ordinal calibration; and the Advisor generates feedback from
both cognitive level and classroom coverage. This staged design keeps
measurement separate from recommendation, reducing the risk that fluent advice
silently changes the diagnosis. It also distinguishes two pedagogically
different moves: broadening participation at the current level versus moving
already-constructive learners toward interaction.

Across real-world STEM and Humanities lectures, MAVE achieves the highest
expert ratings under matched evidence and model inputs. Ablations further show that target reference, multimodal grounding, and
calibrated state estimates each provide distinct value. Under controlled
student-agent simulations, MAVE's rewrites and scaffolds induce more frequent
and deeper ICAP upgrades.

\bibliographystyle{unsrtnat}

\appendix

\end{document}